%% file: weibel.tex
\documentclass[%
 reprint,
 amsmath,amssymb,
 prl,
superscriptaddress
]{revtex4-2}

\usepackage{svg}
\usepackage{graphicx}

\graphicspath{{./figures/}}
\usepackage{amsmath}
\usepackage{siunitx}
\usepackage{mathrsfs}

\usepackage{dcolumn}
\usepackage{bm}
\newcolumntype{L}[1]{>{\raggedright\let\newline\\\arraybackslash\hspace{0pt}}p{#1}}

\usepackage[utf8]{inputenc}
\usepackage[T1]{fontenc}
\usepackage{mathptmx}
\usepackage{lipsum}

\begin{document}

\input{weibel_abridged_nocomments}

\bibliographystyle{apsrev4-2}
\bibliography{ref}

\end{document}

%% file: weibel_abridged_nocomments.tex
\preprint{APS/123-QED}

\title{Observation of electromagnetic filamentary structures produced by the Weibel instability in laser-driven plasmas}

\input{authors.tex}

\date{\today}

\begin{abstract}
We present experimental observations of electron-scale structures in an expanding 
high-energy-density (HED) plasma generated with a modest intensity $\sim$\SI[per-mode = symbol]{2e14}{\watt\per\cm\squared}, $\sim$\SI{1}{\nano\second} laser. 
The observed structures have wavelengths ($\sim$150-220~\si{\micro\meter}) and growth rates ($\sim$0.4-1.0~\si{\per\nano\second}) consistent with an electron-driven Weibel instability where the anisotropy in the electron distribution is small, $A\sim0.002$. 
This instability is found to be a better match to the observed phenomena than other typical field-generation mechanisms found in HED plasmas, including counter-streaming ion Weibel and magnetothermal instabilities.
These observations experimentally demonstrate for the first time that the electron Weibel instability must be considered alongside other magnetic field generation and amplification mechanisms in expanding ablation plasmas, which are ubiquitous in HED research. They also provide physics insight into the generation of magnetic fields in large-scale astrophysical plasmas. 
Additionally, inspection of the magnetic power spectrum shows a possible scaling match to analytic gyrokinetic predictions, $|B_k|^2\propto k^{-16/3}$, at scales below the electron Larmor radius.

\end{abstract}

\maketitle

Magnetic-field-generation mechanisms in plasmas are important in a broad array of fields, from inertial-confinement fusion to astrophysics~\cite{igumenshchev2014self,farmer2017simulation,ckl2009hohlraums,Zweibel}. The Biermann battery is understood to be one of the plausible mechanisms for generating seed magnetic fields in interstellar plasmas; however it is hard to explain how these fields could have been amplified to observed levels, except through subsequent amplification over many orders of magnitude by turbulent dynamo~\cite{Kulsrud2008}. 
The Weibel instability is another candidate for field generation and amplification~\cite{Weibel1959,schlickeiser2003cosmological}.
While saturated Biermann fields scale inversely with the scale length, $B \propto (L_T/d_e)^{-1}$~\cite{Haines1997}, 
the Weibel-generated fields saturate at amplitudes independent of the system size, and thus 
can exceed the Biermann fields by many orders of magnitude for systems with large scale-lengths, as has been predicted by a variety of particle-in-cell (PIC) simulations~\cite{silva2003interpenetrating,medvedev2006cluster,Kulsrud2008,kato2008nonrelativistic,Schoeffler2014,Schoeffler2016,zhou2022spontaneous}.

Theoretical predictions show that anisotropies in the velocity distribution of plasma species are unstable to the Weibel instability~\cite{Weibel1959}. 
The Weibel instability can have different scaling depending on which species is anisotropic, ions or electrons, and the form of the anisotropy~\cite{davidson1972}.
In a bi-Maxwellian electron distribution function (different temperatures in different directions) the degree of temperature anisotropy between directions is linked to both the maximum growth rate $\gamma_{max}$ and the dominant unstable wavenumber $k_{max}$ through separate formulae~\cite{davidson1972}:
\begin{align}
    k_{max} d_e &= \sqrt{A} \label{eq:kmaxA}, \\
    \gamma_{max} \frac{d_e}{v_{th}} &= \sqrt{\frac{8}{27\pi}} A^{3/2}, \label{eq:gammaA}
\end{align}
with the anisotropy $A$ defined as $A=(T_\perp/T_\parallel-1)$,
where $v_{th}$ is the electron thermal velocity, $T_\perp$ and $ T_\parallel$ are the temperatures in the directions perpendicular and parallel to $\vec{k}$, respectively, and $T_\perp>T_\parallel$. 

Previous laser-driven experiments have probed the Weibel instability in different regimes:
fast interpenetrating flows of ions in counter-streaming ablated plasmas from $\sim$\si{\nano\second} long, modest $\sim10^{14}$~\si[per-mode = symbol]{\watt\per\cm\squared} intensity lasers which drive the instability through ion anisotropy (and seek to form ion Weibel instability-mediated "collisionless shocks")~\cite{huntington2015observation,fiuza2020electron,li2019collisionless}, while much higher intensity $\sim$\si{\pico\second}, $10^{19}$~\si[per-mode = symbol]{\watt\per\cm\squared} lasers drive the electron population into a collisionless state with large anisotropies from return currents~\cite{Quinn2012,ruyer2020}. 
In contrast, $10^{14}$~\si[per-mode = symbol]{\watt\per\cm\squared} intensity lasers can drive a single ablation flow devoid of ion counterstreaming, where small electron anisotropies are driven by a number of possible mechanisms, chief of which is the collisions between hot electrons and the essentially free-expanding ion population of the ablating plasma.
These collisions can redistribute electron momentum from the expansion direction to the transverse direction~\cite{thaury2009,thaury2010}.
Although simulations have addressed this free-expansion regime and some experiments~\cite{rygg2008proton,gao2016ultrafast,manuel2013instability} see filamentary structure that they attribute to a number of possible causes, there have been no experiments to our knowledge which systematically investigate the spatial structure and temporal dynamics of the filamentary structures in 
expanding laser-ablated plasmas to the extent where it is possible to draw conclusions on the underlying mechanism. 
These $\sim$\si{\nano\second}-duration, $\sim$\si{\kilo\joule}-energy pulses set up strong density and temperature gradients which support both the Biermann battery (as has been established by prior work~\cite{stamper1971spontaneous,sakagami1979two,nilson2006magnetic,Li2007,Sutcliffe2022biermann}) and, as is demonstrated in this Letter, a thermal anisotropy which supports the electron Weibel instability. 

In this Letter, we present the results of experiments
on the OMEGA laser
investigating the electron Weibel instability in this unexplored regime with intensity $\sim10^{14}$~\si[per-mode = symbol]{\watt\per\cm\squared}. The experiments shed light on the spatial structure and temporal evolution of Weibel-instability-generated magnetic fields, and show properties consistent with
analytical predictions and PIC simulations 
of the electron Weibel instability in expanding plasmas. 

\begin{figure}
    \includegraphics[width=\linewidth]{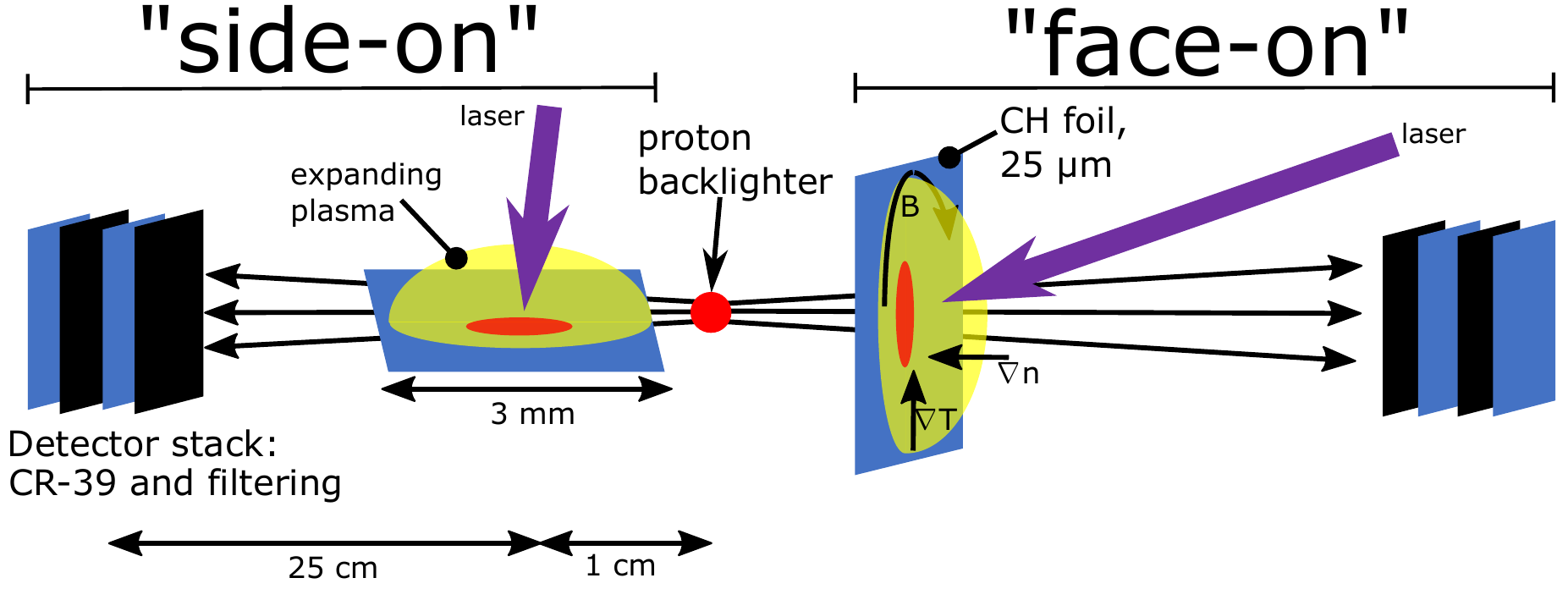}
    \caption{Diagram of the experimental layout. 
    Two experiments are driven simultaneously by a subset of OMEGA laser's beams, while the remainder drive the implosion-based backlighter.
    14.7 MeV protons, the products of D+$^3$He fusion reactions, stream in all directions. Protons that pass through the experimental subject plasmas are deflected by magnetic fields and recorded on CR-39 detectors. }
    \label{fig:experiment}
\end{figure}

Shown in Fig.~\ref{fig:experiment} is a diagram of the experiment. Two 25~\si{\micro\meter} thick polystyrene (CH) foils are driven in identical laser configurations simultaneously, each by 6 beams of the 351 nm OMEGA laser. The pulse is a 1~\si{\tera\watt}, 3~\si{\nano\second} drive (6 beams at 0.5~\si{\tera\watt} and 1~\si{\nano\second} each, staggered in time) 
with a super-Gaussian spot profile with a $1/e$ diameter of 716~\si{\micro\meter}. The laser ablates material from the foil and generates an expanding plasma bubble. Misaligned temperature and density gradients allow the Biermann battery mechanism to generate a large-scale magnetic field oriented circumferentially around the bubble. An implosion-based proton backlighter generates 14.7 MeV protons through the fusion of deuterium with helium-3~\cite{li2006backlighter,li2006prlbacklighter}, which stream in all directions; those that pass through the experiment are individually recorded on a CR-39 detector. The deflections the protons experience in the plasma as a result of magnetic and electric fields manifest as variations in the detected particle flux. 
The experiments are run with two diagnostic configurations: "side-on" with protons probing parallel to the surface of the foil, and "face-on" with  protons  passing through the face of the CH foil. In the face-on configuration the protons are sensitive to the largely perpendicular Biermann magnetic field, while in the side-on view the deflections from the Biermann fields are cancelled out~\cite{petrasso2009doublebubble}. In the side-on view, voids in the recorded proton flux appear near the foil corresponding to large-scale electric fields because the target foil has a net charge. This region is avoided in the reconstruction of fields from the proton radiographs.

Thomson scattering measurements provide a local measurement of electron temperature and density in the expanding plasma bubble. The probe beam and collection optics are focused on a small $\sim$100~$\times$~50~\si{\micro\meter} cylinder offset 1~mm radially and 1~mm vertically from the center of the laser spot on the CH foil. The probe beam is a 45 J, 1 ns pulse of 527 nm light (frequency-doubled from the OMEGA driver's 1053 nm wavelength). 
Light which scatters from the electron plasma wave (EPW) is recorded  with spectral and temporal information resolved~\cite{froula2010plasma}. This data is fit with a model of EPW Thomson scattering spectra to extract key aspects of the plasma conditions: electron temperature and electron density, as a function of time~\cite{froula2010plasma}.
After the very hot front of the bubble passes the measurement volume at $\sim$1~ns, the electron temperature and density are observed to be roughly constant at 900~eV and $6\times10^{19}$ electrons per cubic centimeter.

\begin{figure}
    \includegraphics[width=\linewidth]{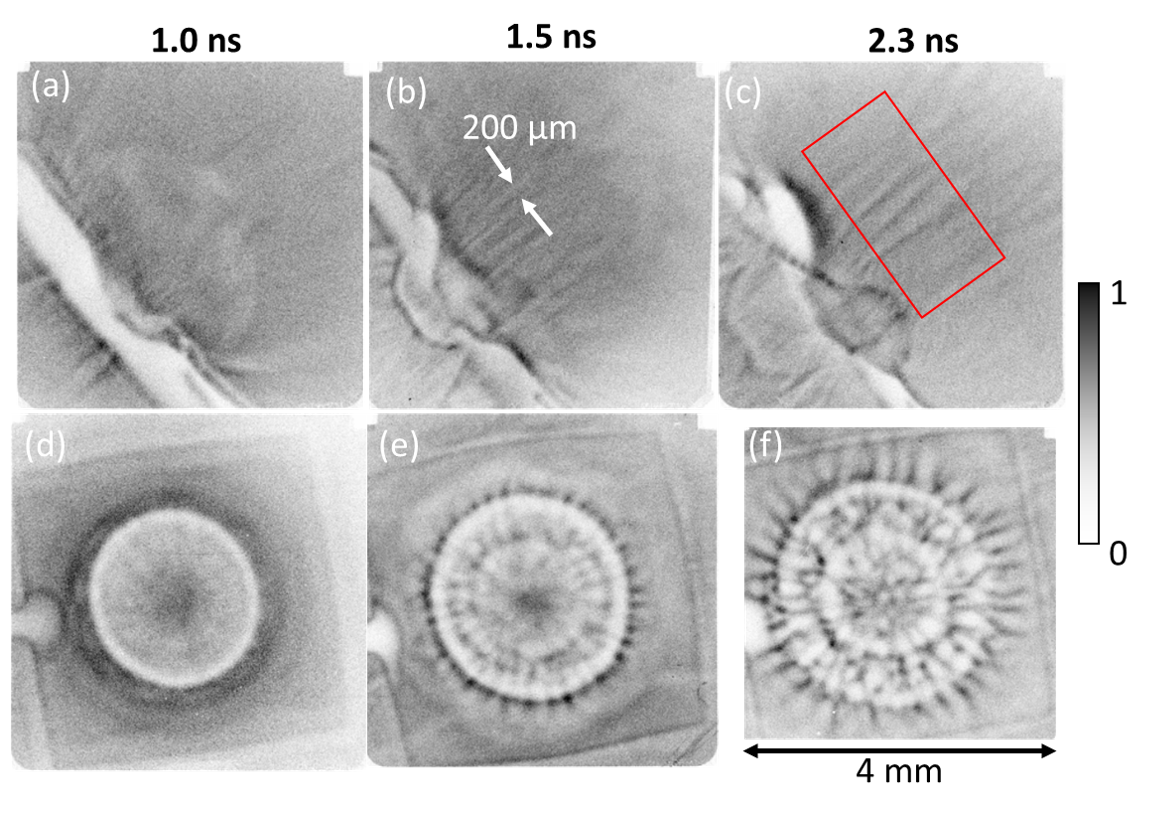}
    \caption{D$^3$He-proton radiographs of side-on (a-c) and face-on (d-f) configurations. The field of view of the radiographs is a $\sim$4~mm side length square. Filaments are of order 200~\si{\micro\meter} at 1.5~\si{\nano\second}, as seen in (b). The red box in (c) shows an example area for which magnetic field reconstructions are conducted in further analysis.} 
    \label{fig:prad}
\end{figure}

Proton radiographs record the spatial structure and temporal evolution of current filaments through the sensitivity of the protons to magnetic and electric fields.
Shown in Fig.~\ref{fig:prad} are the two sets of radiographs.
The time at which the protons probe the subject plasma is controlled by adjusting the proton backlighter laser drive relative to the subject laser drive in repeated experiments.
The time it takes for protons to travel from the backlighter to the subject plasma is accounted for.
The implosion of the backlighter creates a proton source of size 50~\si{\micro\meter} full-width-half-maximum, which is the limiting factor in spatial resolution for the radiographs~\cite{li2006backlighter}. 
Protons are released within a 100 ps window, which sets the temporal resolution of each individual radiograph~\cite{li2006backlighter,li2006prlbacklighter,johnson2021backlighter}.
At
1.0~\si{\nano\second}, there are few filaments visible in the side-on configuration and none visible in face-on configuration (which is more sensitive to the larger Biermann fields~\cite{petrasso2009doublebubble} and subject to scattering in the foil material, which effectively adds 20~\si{\micro\meter} of spatial blur in the subject plasma coordinates). 
The faint outline of the bubble interface is visible, and indicates a bubble front expansion velocity of order 1000 km/s.
Immediately afterwards, at the 1.5~\si{\nano\second}, we see the emergence of $\sim$200 \si{\micro\meter} filaments extending largely perpendicularly from the foil.
The face-on radiographs reveal that these filaments, superimposed upon the Biermann deflection pattern, have some radial component from the center of the plasma bubble.
At 2.3~\si{\nano\second}, the filaments are very prominent and have longer wavelengths. PIC simulations also show this behavior, as longer wavelengths take a longer time to grow~\cite{Schoeffler2014}.
Especially in the side-on configuration, we are sampling a "forest" of filaments: what is seen is the resulting particle flux from the line-integrated deflections experienced along their trajectory. Interpretation of this is simplified by recent simulation work which shows that feature sizes as measured on a radiograph in a similar geometry are directly related to the filament size (and not, for example, the filament spacing)~\cite{huntington2015observation,levesque2019}. 

The strength and dynamics of the measured magnetic fields are inferred from the reconstruction of proton radiographs.
We have used an “optimal transport” algorithm which infers the deflection field, $\vec{\delta}$, that must have been imparted on the probe protons in order to generate the distribution seen on the detector~\cite{sulman2011efficient,bott2017proton}.
The deflection field is converted to the path-integrated perpendicular magnetic field $BL$ through the relation
\begin{equation}
    BL \equiv \int_0^L \hat{z} \times \vec{B} \; dl = \frac{mcv}{q} \vec{\delta},
\end{equation}
where $m$, $v$, and $q$ are the mass, velocity, and charge of the probing proton. 
The value of $L$ is a multiplicative constant along each line of sight and is not important when evaluating the overall shape of spectra of the reconstructed magnetic field. 

\begin{figure}
    \includegraphics[width=\linewidth]{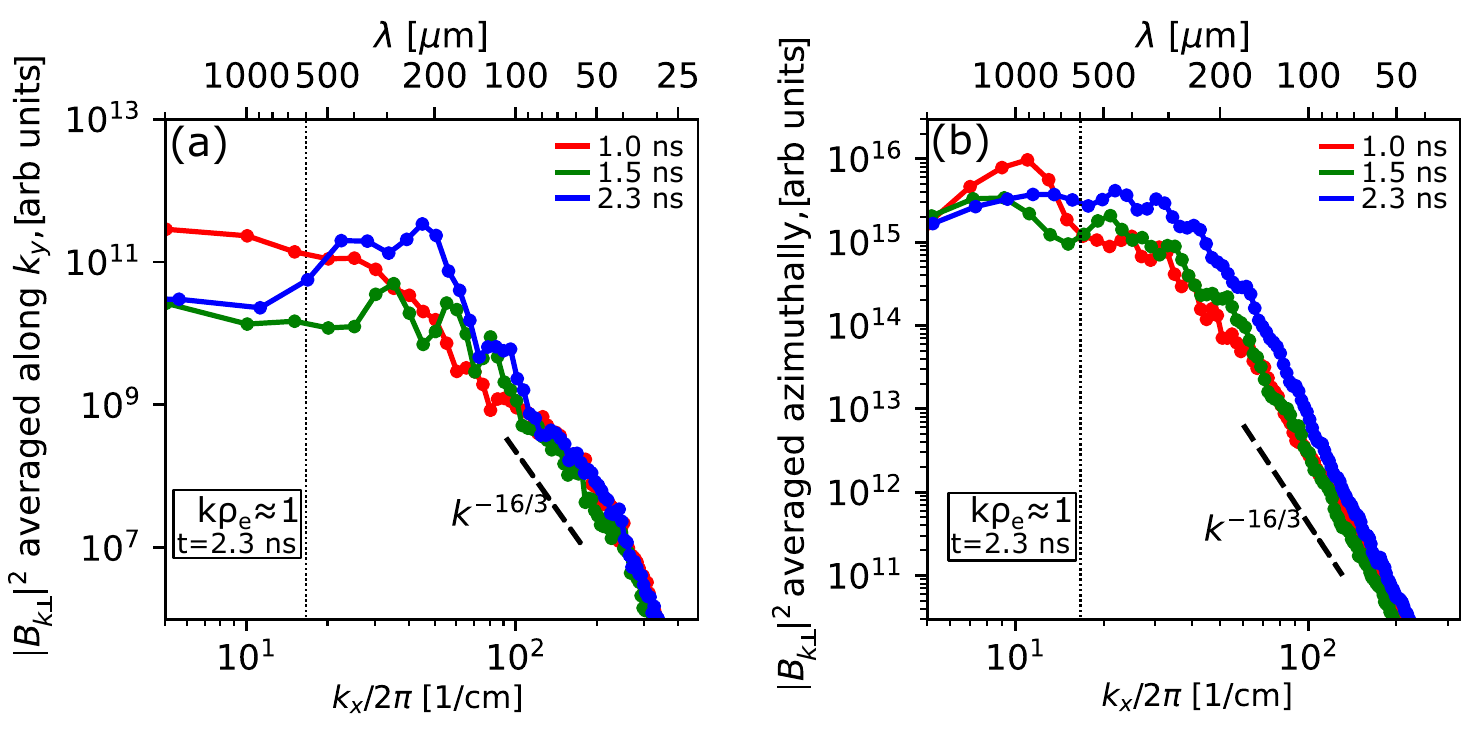}
    \caption{Fourier spectra of the reconstructed magnetic field $|B_{k\perp}|^2$ for the side-on case (a) and the face-on case (b).
    The vertical black dotted line shows the $k$ for which the electron gyroradius is of comparable size to the instability wavelength ($k\rho_e\approx1$) at 2.3~\si{\nano\second}.
    A $k^{-16/3}$ relationship is indicated with the black heavy dashed line for reference.
    The resolution-limited roll-off of the spectrum at 200~cm$^{-1}$ can just be seen in side-on case.}
    \label{fig:Bft}
\end{figure}

Fourier spectra of the reconstructed magnetic fields are computed and plotted in Fig.~\ref{fig:Bft}. 
For early times,
the spectra are dominated by large-scale deflections (i.e. small $k$). 
Spectral content at high wavenumber (associated with filaments) develops with time and grows substantially by the latest recorded time.
The amplitude of the magnetic field, particularly at the wavelengths matching the identified filaments (broadly, 150-220 \si{\micro\meter}), are seen to grow as a function of time.

\begin{figure}
    \includegraphics[width=\linewidth]{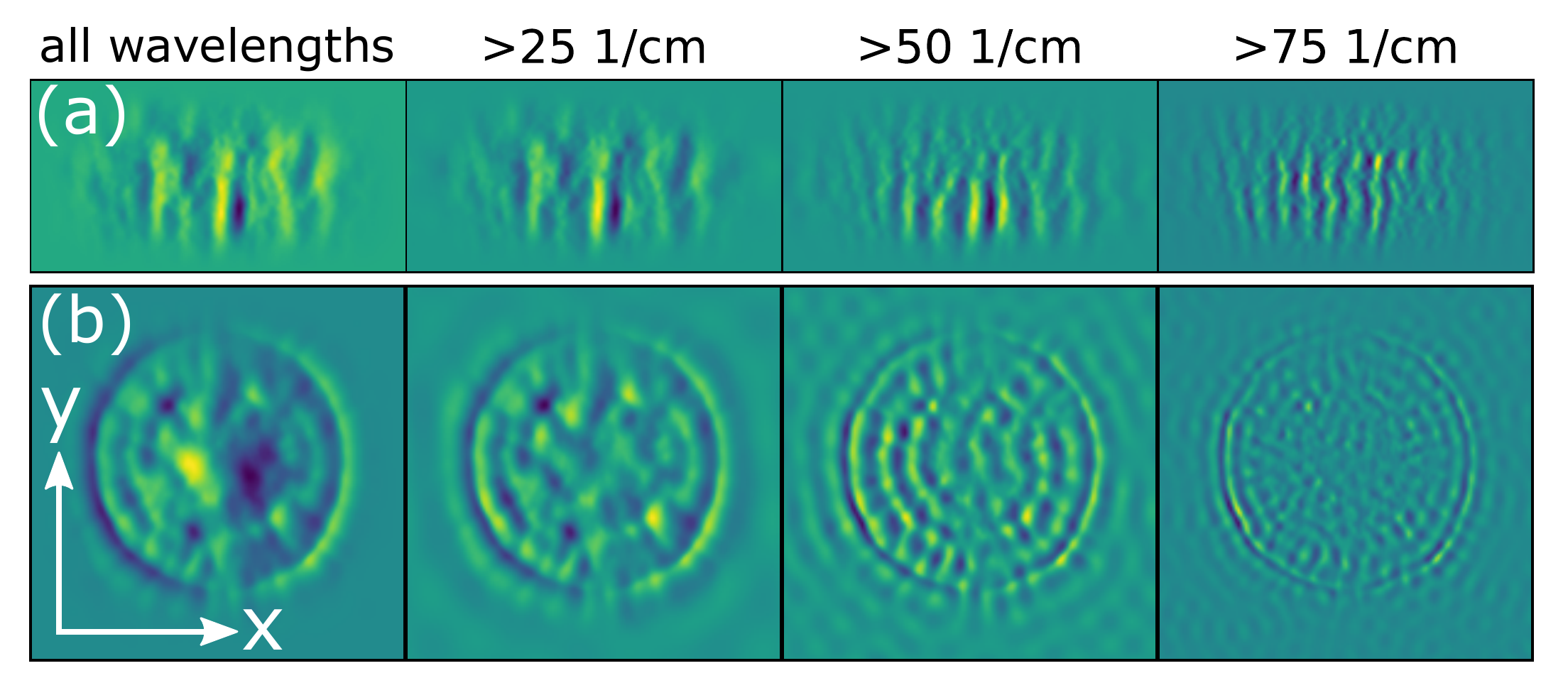}
    \caption{
    Magnetic field reconstruction y-component $B_y$, for the side-on case (a) and the face-on case (b) at 1.5 ns.
    The left column is the raw reconstructed $B_y$, while the other columns show high-passed versions of $B_y$ at $k/2\pi>$25~cm$^{-1}$, 50~cm$^{-1}$, and 75~cm$^{-1}$.
    The colormaps are individually normalized to the largest values in each reconstruction. The dimensions of the side-on reconstructions are approximately 1~$\times$~1.5~mm with some slight variation, while the face-on reconstructions are approximately 3~$\times$~3~mm. 
    }
    \label{fig:bp_summary_small}
\end{figure}

To assess what parts of the spectra are associated with what spatial features, a transverse-component of the reconstructed magnetic field, $B_y$, is high-pass filtered at a variety of wavenumbers $k_{cut}$.
These are plotted
for time = 1.5 ns in
Fig.~\ref{fig:bp_summary_small}, along with the original $B_y$ reconstruction and the orientation of the y-direction (plots for all times can be found in the supplemental material, \cite{supp_bp_summary_small}). 
For side-on radiographs, filamentary structures are present at a variety of wavenumbers.
The face-on reconstructions are more complex: low-wavenumber features contain a great deal of circular Biermann-battery-related spatial features and therefore assessing the correct amplitude of the Weibel fields is difficult. 
The spatial gradients in $B_y$ in face-on reconstructions are sufficiently sharp such that all of the wavenumber bins contain some degree of Biermann-field spatial signature. 

The magnitude of the fields associated with the filaments is quantified by calculating the root-mean-square (RMS) of the y-component of high-passed magnetic field reconstructions, $(BL)_{rms}$, with high-pass cutoff wavenumber $k_{cut}$. 
The growth rate of the magnetic field is calculated using the values $(BL)_{rms}$ as a function of time using the relation for growth at a given $k_{cut}$:
\begin{equation}
    \gamma(t_1 \rightarrow t_2,k_{cut}) = \frac{1}{t_2-t_1} \log \left(\frac{(BL)_{rms}(t_2,k_{cut})}{(BL)_{rms}(t_1,k_{cut})}\right),
\end{equation}
The two measurements in the time intervals $t=[\text{1.0-1.5~\si{\nano\second}}]$ and $t=[\text{1.5-2.3~\si{\nano\second}}]$ are averaged to get the time-averaged growth rate $\gamma(k_{cut})$. 
The quantities $(BL)_{rms}$ and $\gamma(k_{cut})$ are very sensitive to the choice of $k_{cut}$, which is somewhat arbitrary. 
A reasonable high-pass filter cut in $k$ would necessarily include a greater Biermann contribution at early times relative to later times. A sensible compromise seems to be somewhere around 50 cm$^{-1}$, based on inspection of the structures in Fig.~\ref{fig:Bft} and Fig.~\ref{fig:bp_summary_small}.
The calculation of $(BL)_{rms}$ and $\gamma(k_{cut})$ is repeated for a variety of $k_{cut}$'s, and are averaged with a Gaussian weighting function centered at $k_{cut}/2\pi=$50~cm$^{-1}$ with standard deviation 20~cm$^{-1}$.
The result is the Gaussian-weighted lumped growth rate $\gamma$. 
Similarly, the uncertainty in $\gamma$ is the Gaussian-weighted standard deviation of the $\gamma(k_{cut})$ values. 
This process captures our assessment that the scale at which the Weibel features are sufficiently pronounced relative to Biermann features is typically at 50~cm$^{-1}$, but includes the degree of uncertainty and arbitrariness of this assessment as an error in the measured RMS fields $(BL)_{rms}$ and growth rates $\gamma$.
This choice of $k_{cut}$ is the dominant source of error in the resulting $(BL)_{rms}$ and $\gamma$ in the analysis; while experimental details of the proton radiography measurements contribute $\sim$10\% error and the reconstruction (including choices made in preparing the data for reconstruction) contributes no more than 10\% error to measured $(BL)_{rms}$ values, the error from the choice of $k_{cut}$ is found to be larger than this. 

\begin{figure}
    \includegraphics[width=\linewidth]{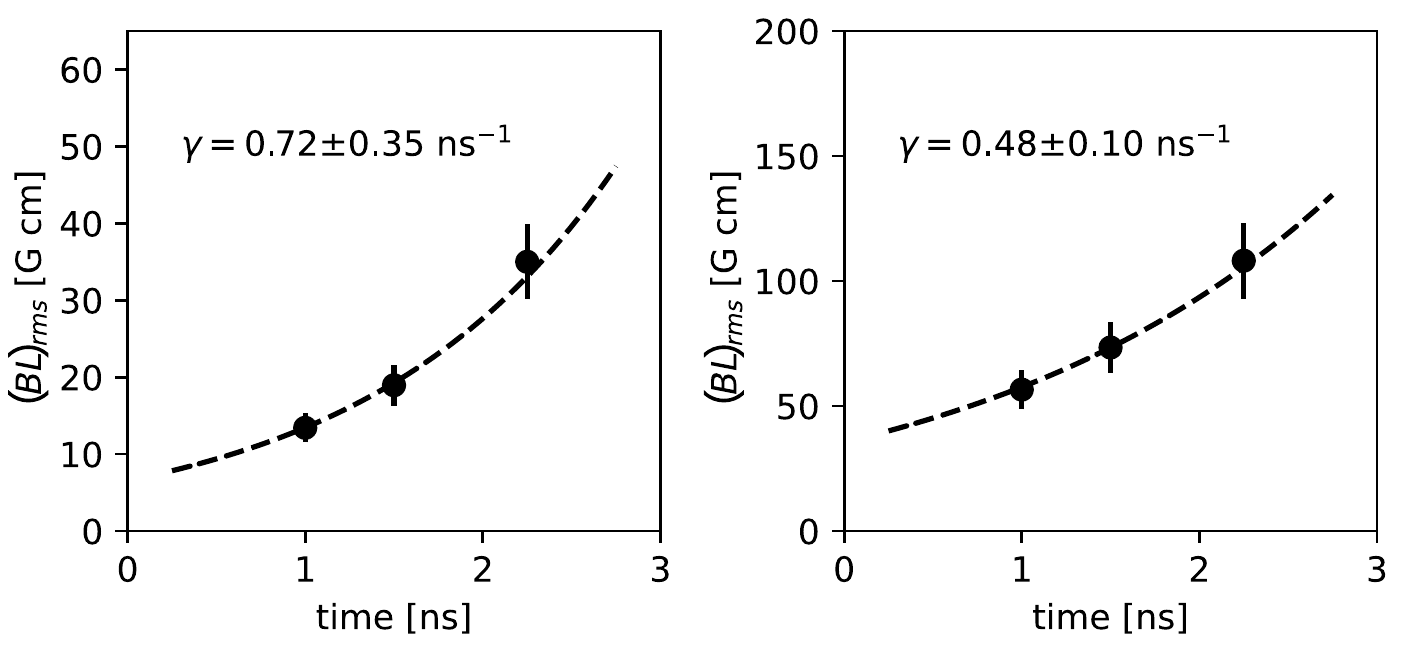}
    \caption{
    Magnetic field amplitude as characterized by calculating the $k$-averaged $(BL)_{rms}$.
    The dashed lines show exponential functions $(BL)_{rms}\propto e^{\gamma t}$ as calculated for each line-of-sight. The errorbars shown are the ``statistical" errorbars, from uncorrelated sources. The errors in $\gamma$ are dominated by the systematic choice of $k_{cut}$.}
    \label{fig:b_vs_t}
\end{figure}

The extracted $(BL)_{rms}$ for each time in the two configurations are plotted in Fig.~\ref{fig:b_vs_t}. 
The error bars shown are the statistical error bars from experimental uncertainties in proton radiography system characteristics and reconstruction uncertainty, a combined 14\% relative error in $(BL)_{rms}$. The $k_{cut}$-related correlated error in $(BL)_{rms}$ is not shown.
The growth rate for the side-on case is found to be $\gamma=$0.72$\pm$0.35~\si{\per\nano\second}, while the face-on case measures $\gamma=$0.48$\pm$0.10~\si{\per\nano\second}. The face-on case likely includes Biermann battery field content in its earlier-time measurements of $(BL)_{rms}$,
thus is expected to underestimate the true growth rate,
explaining the difference seen in the measured~$\gamma$. 

Applying the bi-Maxwellian predicted wavenumber-anisotropy relation, eq.~(\ref{eq:kmaxA}), the measured wavelengths (150-220~\si{\micro\meter}) correspond to an anisotropy of $0.0016\pm^{0.0009}_{0.0005}$ and a growth rate of $0.37\pm^{0.32}_{0.15}$~\si{\per\nano\second}. Conversely, the measured growth rate ($\gamma=$0.72$\pm$0.35~\si{\per\nano\second}) corresponds to an anisotropy of $0.0025\pm0.0008$ (using eq.~(\ref{eq:gammaA})), and a peak wavelength of $148\pm^{36}_{18}$~\si{\micro\meter}. The agreement of these two independent calculations of anisotropy shows that there is an internal consistency in the explanation that an anisotropy in the electron distribution function is driving the formation of current filaments through the Weibel instability. 
Based on the Thomson scattering measurements, electron-ion collisional mean-free-paths are $\sim$50~\si{\micro\meter}, smaller than the plasma system size $\sim$1000~\si{\micro\meter}. As suggested by previous work, interactions between ions and electrons through collisions, as well as acceleration/deceleration by ambipolar electric fields at large pressure gradient interfaces, can serve to generate anisotropy in the electron distribution~\cite{thaury2009,thaury2010}.

Other instabilities and mechanisms are considered and ruled out, as summarized in Table \ref{tab:mechanisms}. First, the ion Weibel instability
involves a hypothetical background plasma into which the ablation flow expands, or similarly a pre-plasma created by early energy delivered by the laser.
Work that has addressed relevant conditions~\cite{matteucci2019weibel} predicts a requisite pre-plasma density of at least $\sim10^{17}$~\si{\centi\meter\tothe{-3}} to achieve growing ion Weibel filaments. The OMEGA chamber has a number density on the order $\sim10^{12}$~\si{\centi\meter\tothe{-3}}, too small to contribute. Similarly, the OMEGA laser's temporal contrast is excellent and we utilized a well-tested pulse shape. Moreover, Ref.~\cite{matteucci2019weibel} shows the morphology of the growing filamentary fields, which occur at the interface of the ablation plume and background plasmas, as opposed to throughout the ablation plume as we observe in our radiographs (Fig.~\ref{fig:prad}). 
Additionally, a formula from PIC simulations which predicts the onset time of Weibel instability suggests that the system should be stable to the ion Weibel instability until at least $t\sim$\SI{40}{\nano\second}~\cite{thaury2010}.

By using the formulae for maximum growth rate and wavelength of maximum growth rate of the magnetothermal instability (MTI) found in Ref.~\cite{manuel2013instability}, applying the parameters measured by Thomson scattering, taking thermal and density gradients to be approximately \SI{750}{\micro\meter} (consistent with laser profile and Thomson measurements), and taking $Z=$3.5 (fully ionized 1:1 CH plasma),
the peak growth wavelength is predicted to be $\sim$190-\SI{250}{\micro\meter}, similar to the observed structures. The growth rate, however, is substantially higher than what is observed: estimates put the lower bound on growth rate at \SI{10}{\per\nano\second}.
The growth rate being so fast relative to experimental time scales suggests that if the MTI were indeed present at the scales observed it should be able to dominate magnetic field production, but this is not seen.
The saturation condition for MTI is the same as that for the electron Weibel instability: when the cyclotron radius of the electrons from the perturbed (instability-related) magnetic field becomes small enough to be on the scale of the transverse wavelength of the instability, $k\rho_{e} \approx 1$.
This is not achieved at any time for wavelengths associated with the filaments; the wavelength at which this condition holds for the final observed time in both side-on and face-on measurement configurations is $\sim$\SI{600}{\um} (as seen in Fig. \ref{fig:Bft}), much larger than the scale
of the observed filaments.

Lastly, the electrothermal instability (ETI) is a potential source of magnetic field in laser-driven plasmas, but is restricted to regions where the mean free path of electrons is much smaller than the electron inertial length~\cite{Haines1997}.
In the expanding ablation plasma the electron inertial length is $\sim$1~\si{\micro\meter}, much smaller than the $\sim$50~\si{\micro\meter} mean free path.
The ETI is more significant in generating magnetic field closer to the ablation region, where densities are much higher and temperatures fall off towards the dense and cold bulk of the foil. 

\begin{table}[t]
\noindent
\begin{tabular}{ |l|L{0.25\columnwidth}|L{0.25\columnwidth}| } 
 \hline
 Observation/Mechanism &
 Growth Rate (1/ns) &
 Wavelength (\si{\micro\meter}) \\ 
 \hline\hline
  Observed filaments &
  $\sim$0.4-1.0 &
  $\sim$150-220 \\ 
 \hline
  Electron Weibel (with A=0.002) &
  $\sim$0.5 &
  $\sim$170 \\ 
 \hline
 Ion Weibel &
  \multicolumn{2}{L{0.50\columnwidth}|}{Relevant PIC simulations~\cite{matteucci2019weibel} show poor spatial and  temporal match to data.} \\ 
 \hline
  Magnetothermal instability &
  $\gtrapprox$10 &
  $\sim$190-250 \\ 
 \hline
  Electrothermal instability &
  $\sim$0.04 &
  Very large \\
  \hline
\end{tabular}
\caption{\label{tab:mechanisms} Observed parameters and parameter estimates for candidate magnetic field generation mechanisms in laser-generated plasmas. The growth rate and wavelengths of candidate mechanisms are estimated from dispersion relations in the literature using the Thomson scattering-measured typical electron temperature (\SI{900}{\electronvolt}) and density (\SI{6e19}{\centi\meter\tothe{-3}}) in the ablation plasma plume.}
\end{table}

A feature is seen 
at high wavenumbers which is of separate interest. The slope of the power spectrum
at wavenumbers
greater than the main Weibel-related peaks but below the resolution limit $k/2\pi\sim$200~cm$^{-1}$ (corresponding to a $\sim$50~\si{\micro\meter} imaging resolution dictated by the backlighter source size) tends to a constant power law relation, $|B_k|^2 \propto k^{-16/3}$, as indicated in Fig.~\ref{fig:Bft}. 
This scaling was predicted in gyrokinetic analysis of magnetized turbulence at sub-electron scales~\cite{schekochihin2009astrophysical} and subsequently observed in 3D PIC simulations of laser-target interactions in a configuration that reasonably matches that of our experiment~\cite{Schoeffler2014,Schoeffler2016}. This could be a promising subject of future investigation.

To conclude, electron-Weibel-instability-generated magnetic-field structures and dynamics have been measured and quantified in a laser-produced, expanding plasma for the first time. Thomson scattering is used to accurately characterize the plasma temperature and density. Time-resolved proton radiography reveals the structure and temporal evolution of the filaments. Reconstruction of the magnetic fields from radiographs quantifies the magnetic field and its growth. We have shown that the observed structures have a wavelength and growth rate consistent with PIC predictions of the electron Weibel instability~\cite{Schoeffler2014,Schoeffler2016}, validating this prior work. 
Another similarity between our observations and the PIC results is the theoretically-predicted $|B_k|^2 \propto k^{-16/3}$~\cite{schekochihin2009astrophysical} spectral shape at larger wavenumbers than the Weibel-related peaks, which prompts further investigation.
Quantitative and qualitative arguments exclude the role of other instabilities in forming the measured filamentary structure and dynamics.
This work demonstrates
that the electron Weibel instability must be considered alongside other magnetic field generation and amplification mechanisms in expanding ablation plasmas,
and legitimizes scenarios for magnetic field generation in astrophysical plasmas based on the electron Weibel instability. 

This work was supported in part by U.S. Department of Energy NNSA MIT Center-of-Excellence (DE-NA0003868), NLUF (DE-NA0003938), and NSF CAREER award No. 1654168.

%% file: authors.tex


\def \mit {Plasma Science and Fusion Center, Massachusetts Institute of Technology, Cambridge, Massachusetts 02139, USA}
\def \llnl {Lawrence Livermore National Laboratory, Livermore, California 94550, USA}
\def \princeton {Princeton University, Princeton, NJ}
\def \lle {Laboratory for Laser Energetics, Rochester, NY}

\author{G.D.~Sutcliffe}
\email[Corresponding author's email: ]{gdsut@mit.edu}
\author{P.J.~Adrian}
\author{J.A.~Pearcy}
\author{T.M.~Johnson}
\author{J.~Kunimune}
\affiliation{\mit}

\author{B.~Pollock}
\author{J.D.~Moody}
\affiliation{\llnl}


\author{N.F.~Loureiro}
\author{C.K.~Li}
\affiliation{\mit}
